\documentclass[11pt]{article}
\usepackage[margin=1in]{geometry}
\usepackage{graphicx}
\usepackage{subcaption}
\usepackage{booktabs}
\usepackage{amsmath}
\usepackage{float}
\usepackage[section]{placeins}
\usepackage{hyperref}
\usepackage{microtype}
\renewcommand{\thefootnote}{\fnsymbol{footnote}}

\title{Quantifying Officiating Impact in the NBA:\\A Referee Impact Metric Analysis Using ESPN Win-Probability Data}
\author{
\begin{tabular}{cc}
Nirek Duma\footnotemark[1] & Leo Benaharon\footnotemark[1]\\
\texttt{nireksai@gmail.com} & \texttt{lbenahar@purdue.edu}\\
\multicolumn{2}{c}{Purdue University}
\end{tabular}
}
\date{}

\begin{document}
\maketitle
\footnotetext[1]{Equal contribution.}
\renewcommand{\thefootnote}{\arabic{footnote}}
\setcounter{footnote}{0}

\begin{abstract}
Over the past century, basketball analytics has moved from simple box-score rates toward complex context-aware measures that evaluate events by their expected effect on game outcomes. Officiating analysis has not made the same transition: existing work and public discussion still rely heavily on foul rates, foul differentials, reviewed late-game correctness labels, or team/player benefit from calls. This leaves an empirical gap because a low-leverage foul in a decided game should not be treated as equivalent to a whistle that materially shifts win probability in a close game. To address this gap, we introduce the Ref Impact Metric (RIM), a game-level statistic that aggregates the absolute win-probability movement attached to foul events, measuring the impact of each referee for each game. Using ESPN game-summary and win-probability data for NBA seasons 2021-2022 through 2024-2025, we show that RIM is empirically distinct from both foul volume and foul disparity, identify regular-season and postseason referee distributions, and examine home/away, team-side, and referee-team heterogeneity. We then use linear controls intentionally as stress tests: conditioning on home status, team, opponent, season, and postseason series state asks which descriptive outliers persist after basic contextual adjustment. The results show that several team-side and referee-team patterns remain visible after conditioning, but omitted-variable robustness diagnostics indicate that these patterns should be interpreted as observational screening signals rather than evidence of intent, misconduct, or whistle-level responsibility by any single official. Our contribution to the literature is foundational, and we emphasize that this framework should be tested with different win probability models and further causal inference.
\end{abstract}

\section{Background}
The effort to quantify individual impact in team sports far predates modern tracking systems or the popularization of the ``Moneyball'' era \cite{lewis2003}. The historical development of advanced player statistics has generally followed a distinct methodological progression through several analytical eras. The initial ``rate-stat era'' focused on correcting raw counting statistics for playing time and opportunity. This framework was established early across major sports, dating back at least 159 years to Henry Chadwick's 1867 proposal for safe-hit and total-base averages in baseball \cite{schwarz2004}. Basketball followed with the introduction of field-goal percentage in the 1946--47 BAA season \cite{basketballReference1947}. For decades, player and team evaluation relied heavily on these traditional box-score metrics, summarizing impact through per-game averages of points, rebounds, assists, and fouls.

Analysts eventually recognized that raw aggregates and simple rates failed to account for game context, prompting a shift into the ``efficiency era.'' Pioneered in the late 1990s and early 2000s by analysts such as Dean Oliver, basketball moved toward possession-based frameworks \cite{oliver2004,kubatko2007}. Metrics like offensive and defensive rating evaluated teams and players per 100 possessions, showing that raw counting statistics were heavily shaped by pace \cite{oliver2004,kubatko2007}. This era established a central analytical principle: volume without the context of opportunity is inherently misleading.

Over the last decade-plus, NBA analytics has transitioned into the ``deep context era.'' Following the league-wide implementation of optical tracking cameras in the 2013--14 season, evaluation shifted from box-score outputs toward situational expected values \cite{cervone2014,cervone2016}. Modern basketball analytics now routinely weights events by game state, using measures such as shot quality, expected effective field-goal percentage based on defender proximity, and win probability added \cite{cervone2014,cervone2016,deshpande2016}. In this era, events are judged not only by whether they occurred, but by their asymmetric leverage on the game's outcome.

While player and team evaluation has transitioned into the context era, public and academic evaluation of officiating has largely remained closer to the rate-stat era. Much of the existing literature and media discourse still relies heavily on raw foul counts, normalized whistle rates, or simple free-throw disparities. Even rigorous recent studies on officiating bias and accuracy primarily use reviewed-call datasets, such as the NBA's Last Two Minute Reports, to evaluate discrete decision accuracy rather than the dynamic value of those decisions for the game state itself \cite{nbaL2M}.

This creates a fundamental methodological gap. Just as a low-leverage jump shot in a blowout does not carry the same expected win-probability impact as a tied game-winner, a foul called in the first quarter does not carry the same mathematical weight as a foul called in the final minute of a one-possession playoff game. If the analytical goal is to quantify officiating impact rather than mere officiating volume, the evaluation framework must transition into the context era. To measure how referees influence expected outcomes, foul events must carry situational weight.

\section{Introduction to RIM}
Prior NBA officiating research has generally relied on foul rates, reviewed-call labels, or call-benefit measures. Price and Wolfers use personal fouls per 48 minutes to study racial matching between players and referees, and Pope, Price, and Wolfers revisit that setting after public scrutiny \cite{price2010,pope2018}. Price, Remer, and Stone examine shooting fouls, non-shooting fouls, turnovers, home-team effects, and playoff-series incentives \cite{price2012}. Deutscher, Gong, Pelechrinis, and Mocan and Osborne-Christenson use Last Two Minute Report decisions, including correct calls, incorrect calls, correct non-calls, and incorrect non-calls, to study home bias, crowd effects, racial bias, and team/player benefit from calls \cite{nbaL2M,deutscher2015,gong2022,pelechrinis2023,mocan2024}.

To bridge this methodological gap, this paper introduces the Ref Impact Metric (RIM), a win-probability-based measure of foul-related officiating impact. Applying this metric to ESPN play-by-play data from the 2021--22 through 2024--25 NBA seasons \cite{espnNBA}, we characterize the distribution of officiating leverage to separate high-volume referees from high-leverage referees. Rather than attempting strict causal identification, we use cluster-robust linear regressions and omitted-variable sensitivity diagnostics to distinguish descriptive anomalies from patterns that persist after conditioning on team, opponent, and playoff series state. The primary contribution of this work is to provide a reproducible framework that brings officiating evaluation into the context era.

\subsection{Operational Context and Data Limitations}
Transitioning from theoretical leverage frameworks to empirical measurement requires navigating the structural realities of NBA data. During the sample period, the league employed a rotating roster of roughly 65 to 75 officials each season, deployed across 1,230 regular-season games in three-person crews. While these officials make hundreds of split-second decisions per game, public play-by-play feeds record only the assigned crew for the game and do not reliably attribute individual foul calls to the specific referee who blew the whistle.

Consequently, any analysis relying on these public feeds must be interpreted as a crew-level association rather than individual attribution. Within the RIM framework, when a referee is identified as a statistical outlier, it indicates that games involving their assigned crew experience unusual win-probability variance. It does not certify that the specific individual is single-handedly generating the leverage. Referee assignment is also not random: more experienced or higher-profile officials may be assigned to games that are expected to be more competitive or higher stakes. A high RIM profile can therefore reflect the types of games assigned to a referee as well as anything about the referee's own officiating environment. Acknowledging these limitations is critical: RIM acts as a macro-level diagnostic tool to identify high-leverage referee-game associations, which can then direct targeted, granular review using internal tracking or optical data.

\subsection{Interpreting RIM as Impact Rather Than Bias}
Throughout the paper, RIM should be read as an impact metric rather than a bias metric. It measures the amount and direction of win-probability movement attached to foul events in games involving a referee's assigned crew. It does not, by itself, determine whether a call was correct, whether a team was unfairly advantaged, or whether any referee acted with bias or intent.

This distinction is especially important for signed team RIM and referee-team pairings. Signed team RIM can show which team benefited in expected-outcome terms from the observed sequence of foul events, but directional movement should not be equated with bias. A positive signed RIM means that foul events coincided with movement toward a team, not that the movement was caused by incorrect, preferential, or biased officiating. Similarly, referee-team outliers describe unusual impact patterns relative to additive baselines. They are screening statistics for identifying games or pairings that may warrant closer review, not standalone evidence of biased treatment.

\section{Metric Definition and Validation}
\subsection{Data and notation}
The data come from ESPN game summaries and win-probability feeds for the NBA seasons 2021--22 through 2024--25 \cite{espnNBA}. The pipeline records foul events, crew assignments, game context, and win-probability samples immediately before and after each foul. ESPN identifies the officiating crew assigned to a game, but it does not assign each whistle to one official. The unit of analysis is therefore the crew-game or referee-game association, not whistle-by-whistle attribution to a specific referee. Any referee-level result should be read as a pattern in games involving that official's assigned crew. Across the four seasons of referee-team data, there were 4,876 regular-season games and 356 postseason games.

For foul event \(e\) in game \(g\), let \(w^-_{eg}\) and \(w^+_{eg}\) be the win probabilities immediately before and after the event. Define event leverage as
\[
\ell_{eg}=\left|w^+_{eg}-w^-_{eg}\right|.
\]
If game \(g\) contains \(n_g\) foul events, the game-level RIM statistic is denoted by
\[
r_g=\sum_{e=1}^{n_g}\ell_{eg}.
\]
The average swing per call is
\[
c_g=\frac{r_g}{n_g}.
\]
Thus \(r_g\) measures total foul-related win-probability movement, while \(c_g\) isolates leverage per whistle.

For team-level analyses, let \(f_{ig}\) be the foul count charged to team \(i\), and let \(o_{ig}\) be the foul count charged to its opponent. Signed foul disparity is
\[
s_{ig}=o_{ig}-f_{ig}.
\]
Positive values favor team \(i\), because they indicate that its opponent was charged with more fouls. Let \(u^-_{ieg}\) and \(u^+_{ieg}\) be team \(i\)'s win probability immediately before and after foul event \(e\). Signed team RIM is
\[
q_{ig}=\sum_{e=1}^{n_g}\left(u^+_{ieg}-u^-_{ieg}\right).
\]
Positive values indicate that the observed sequence of foul events coincided with net win-probability movement toward team \(i\). Negative values indicate net movement against team \(i\).

\subsection{Component validity checks}
\begin{figure}[!htbp]
    \centering
    \includegraphics[width=0.84\linewidth]{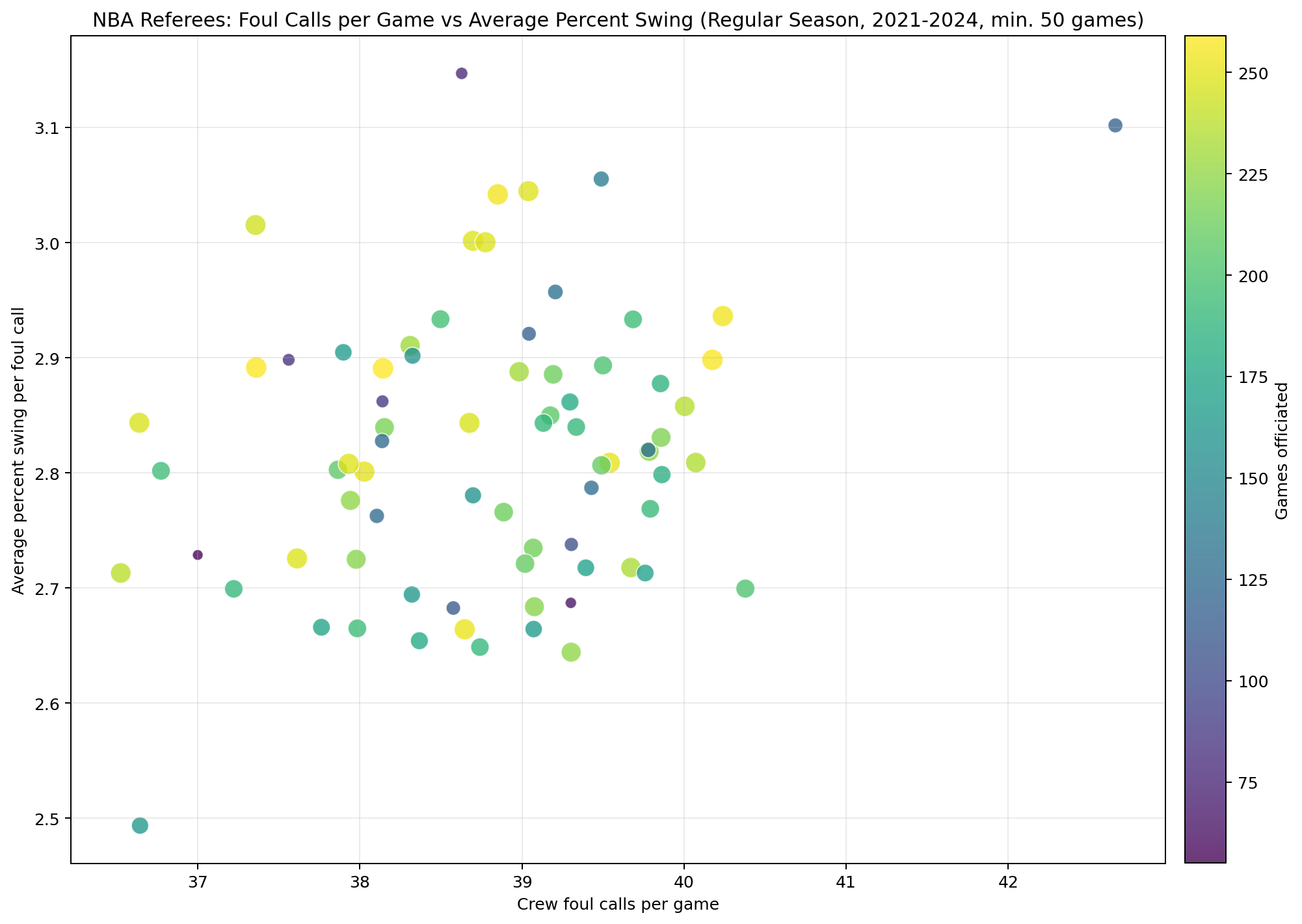}
    \caption{Regular-season foul calls per game plotted against average percent swing per call for referees with at least 50 games from the 2021--22 through 2024--25 seasons. The weak association supports treating volume and leverage as distinct inputs to RIM.}
    \label{fig:calls-vs-swing}
\end{figure}

Before using RIM substantively, the metric should pass a basic validity check: its inputs should not collapse into a single mechanical dimension. Figure~\ref{fig:calls-vs-swing} plots average foul calls per game against average percent swing per call among regular-season referees with at least 50 games. The relationship is weak, indicating that whistle volume and whistle leverage vary separately across referee-game samples.

\begin{figure}[!htbp]
    \centering
    \includegraphics[width=0.84\linewidth]{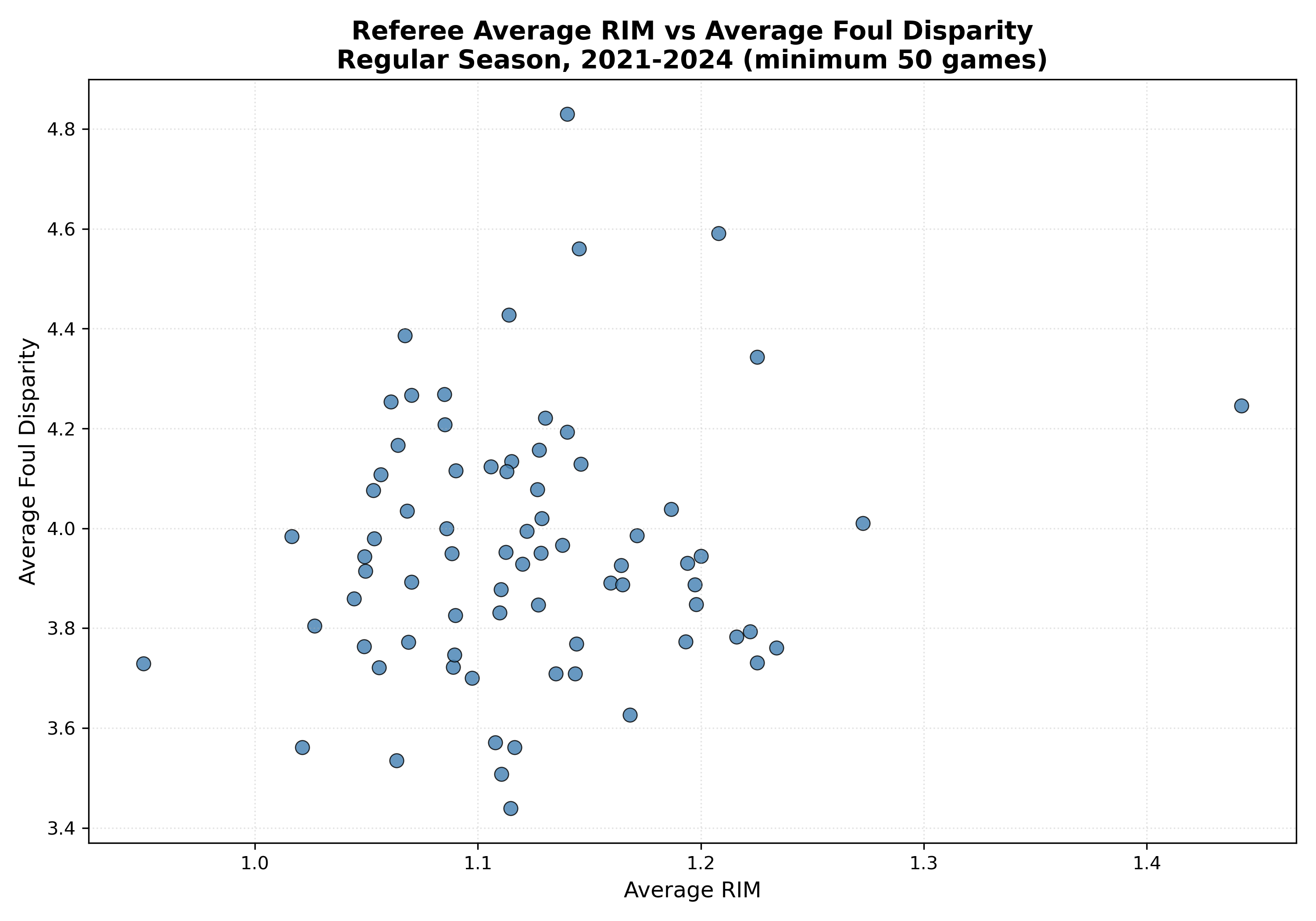}
    \caption{Regular-season average RIM plotted against average foul disparity for referees with at least 50 games. The two measures capture related but empirically distinct features of officiating.}
    \label{fig:rim-vs-disparity}
\end{figure}

Figure~\ref{fig:rim-vs-disparity} compares average RIM with average foul disparity, a more conventional measure in officiating discussions. The two quantities are not strongly correlated. This distinction is important: foul disparity measures imbalance in foul counts, while RIM measures the outcome leverage attached to foul events.

Together, the two checks support the metric's face validity as a distinct descriptive object. They show that RIM is not reducible to whistle volume or foul-count imbalance alone. The metric should therefore be read as a leverage-weighted screening measure: it identifies where foul events carried unusually large win-probability movement, then motivates closer review of the game context behind those patterns.

\FloatBarrier

\section{Regular-Season RIM Patterns}
\begin{figure}[!htbp]
    \centering
    \includegraphics[width=0.98\linewidth]{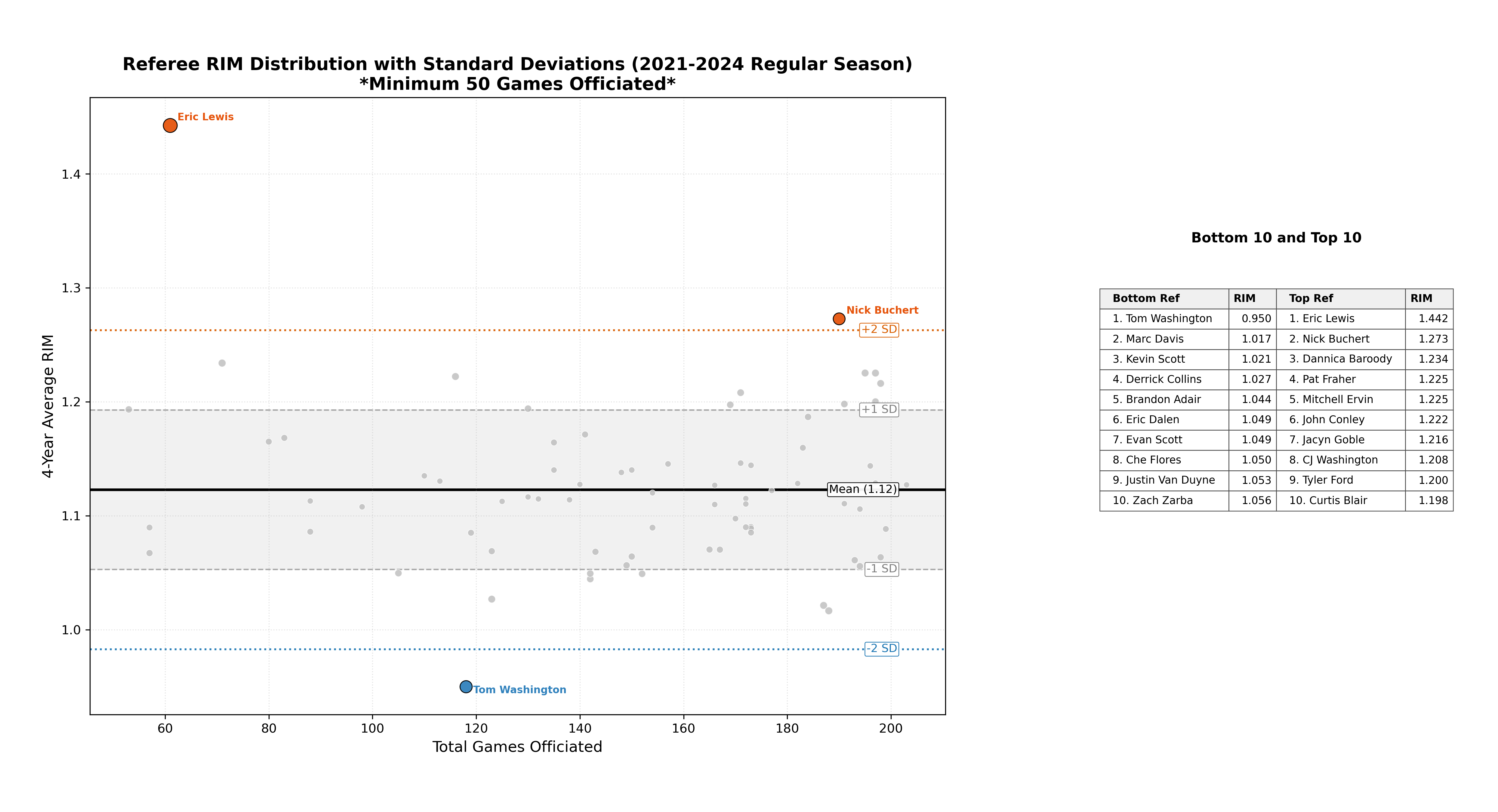}
    \caption{Regular-season RIM distribution with a one-standard-deviation band and a side table of the bottom ten and top ten referees by average RIM.}
    \label{fig:regular-rim-scatter}
\end{figure}

Figure~\ref{fig:regular-rim-scatter} displays the regular-season referee distribution. Most officials cluster near the mean and inside the one-standard-deviation band, but a small number of referees occupy the tails. Eric Lewis and Nick Buchert appear on the high end, while Tom Washington appears on the low end. Because the plot also shows games officiated, high average impact is not simply a function of having a larger sample of games.

\begin{figure}[!htbp]
    \centering
    \includegraphics[width=0.78\linewidth]{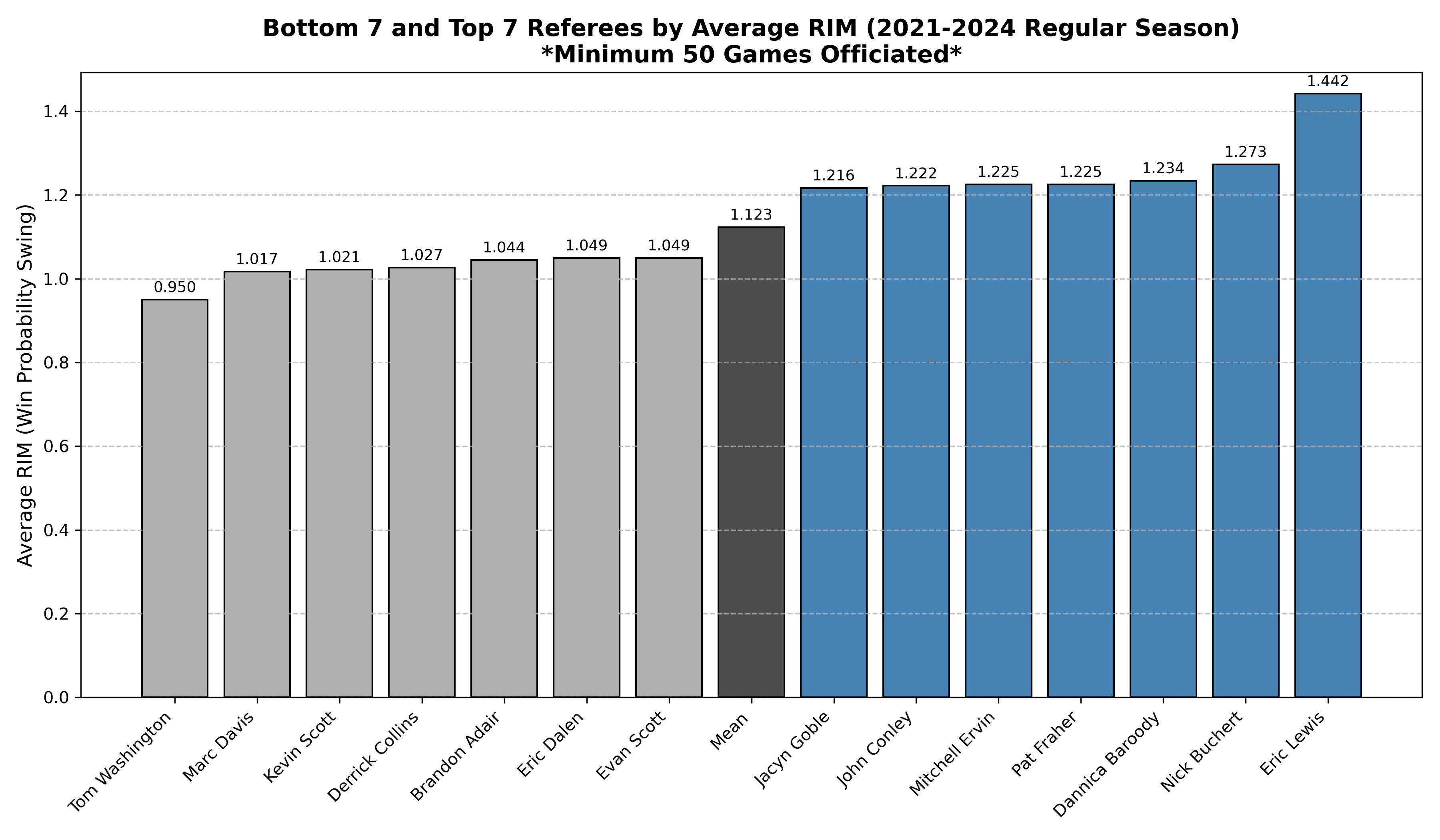}
    \caption{Bottom seven, mean, and top seven regular-season referees by average RIM, using a 50-game minimum.}
    \label{fig:top-rim}
\end{figure}

Figure~\ref{fig:top-rim} gives a more direct comparison of the lower and upper tails. The bottom-seven, mean, and top-seven format shows that average RIM varies meaningfully across qualified referees. This variation motivates the decomposition in the next figure: high average RIM could reflect more whistles, higher leverage per whistle, or both.

\begin{figure}[!htbp]
    \centering
    \begin{subfigure}[t]{0.49\textwidth}
        \centering
        \includegraphics[width=\linewidth]{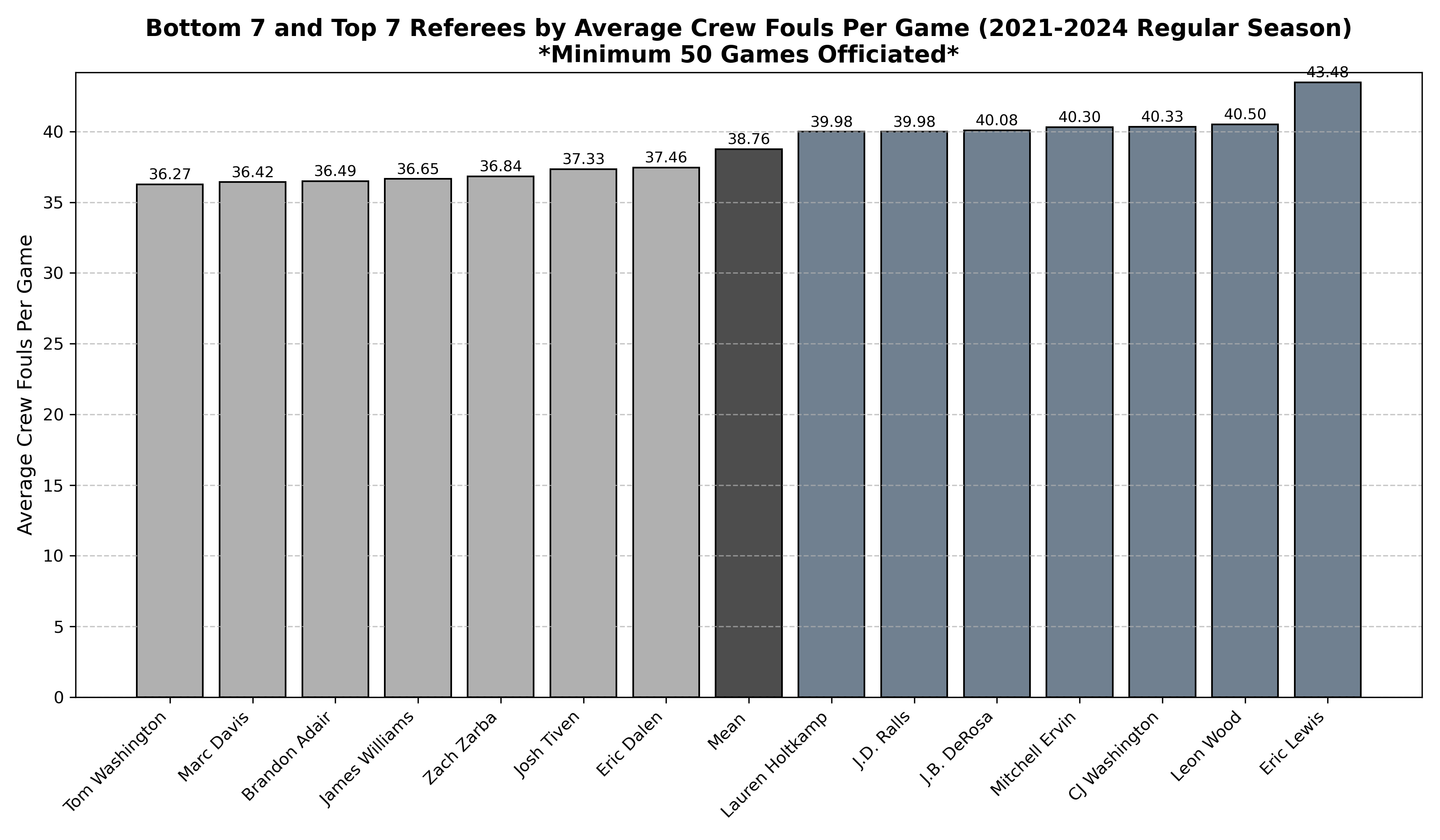}
        \caption{Average crew fouls per game.}
    \end{subfigure}
    \hfill
    \begin{subfigure}[t]{0.49\textwidth}
        \centering
        \includegraphics[width=\linewidth]{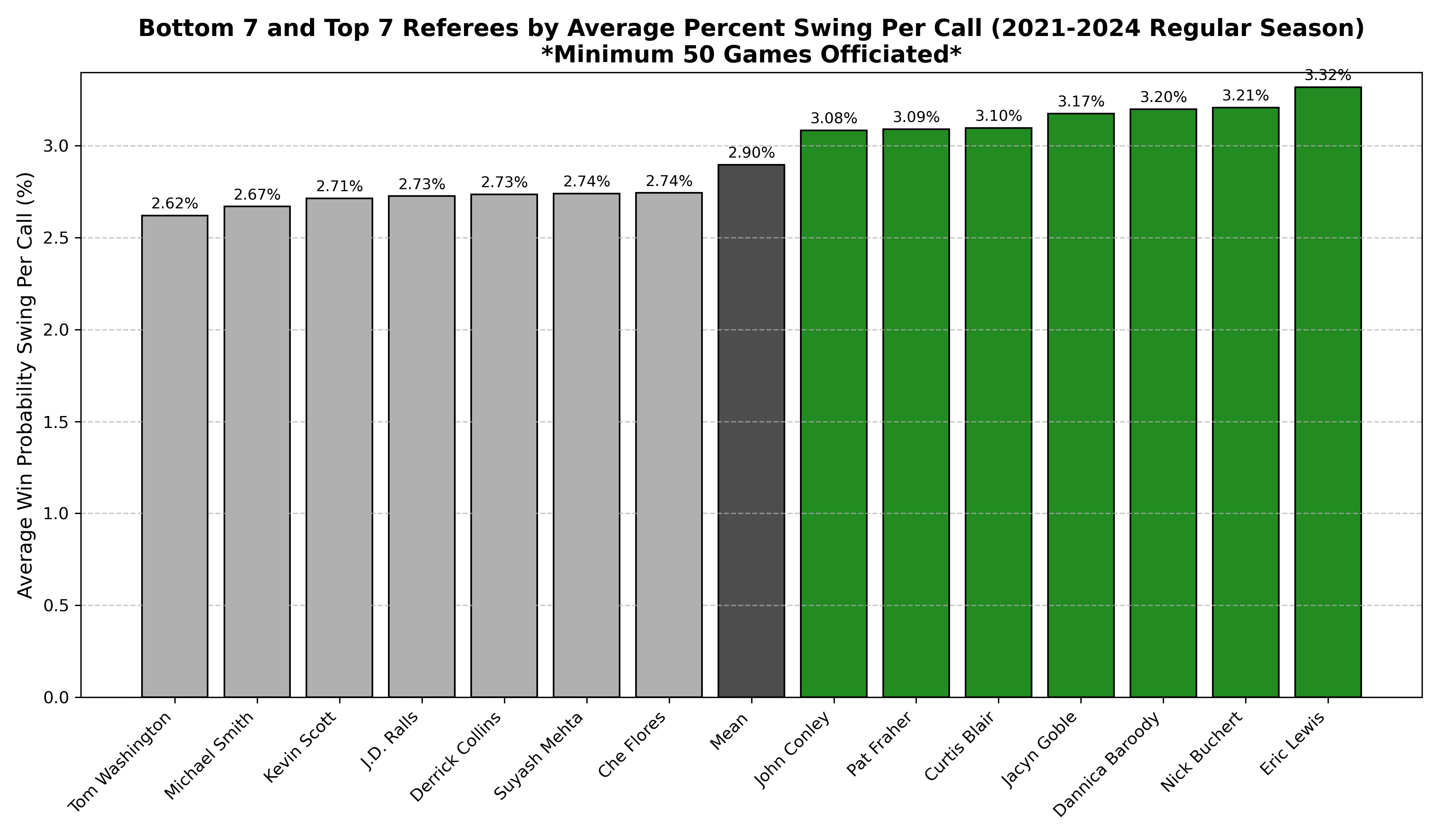}
        \caption{Average percent swing per call.}
    \end{subfigure}
    \caption{Volume and leverage jointly explain why some referees have high or low average RIM.}
    \label{fig:volume-and-swing}
\end{figure}

Figure~\ref{fig:volume-and-swing} separates volume from leverage. The foul-volume panel identifies crews associated with more calls per game, while the swing-per-call panel identifies officials whose foul events carry more win-probability movement per whistle. The rankings overlap but are not identical, reinforcing that RIM combines two substantively different dimensions.

\begin{figure}[!htbp]
    \centering
    \includegraphics[width=0.94\linewidth]{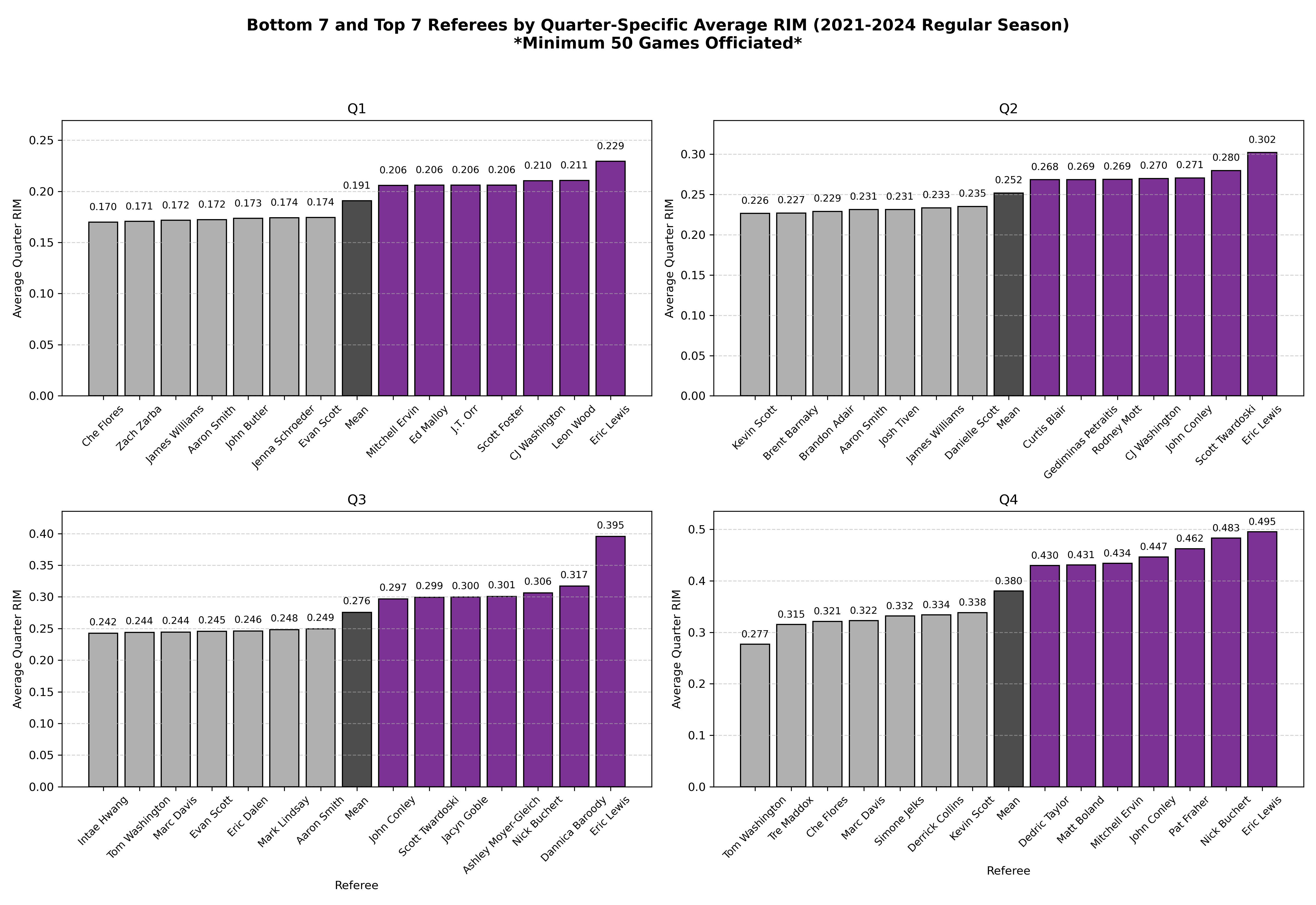}
    \caption{Quarter-specific average RIM by referee. Eric Lewis remains elevated across all four quarters, so his season-level profile is not explained solely by fourth-quarter foul volume.}
    \label{fig:quarter-rim}
\end{figure}

Eric Lewis is high in average RIM, swing percentage, and foul volume, so a natural concern is that the result may be driven by late-game foul counts. Figure~\ref{fig:quarter-rim} weakens that explanation. His elevated RIM profile appears across quarters rather than only in the fourth quarter, which is more consistent with a broader leverage pattern than a purely endgame pattern.

\FloatBarrier

\section{Postseason RIM Patterns}
\begin{figure}[!htbp]
    \centering
    \includegraphics[width=0.84\linewidth]{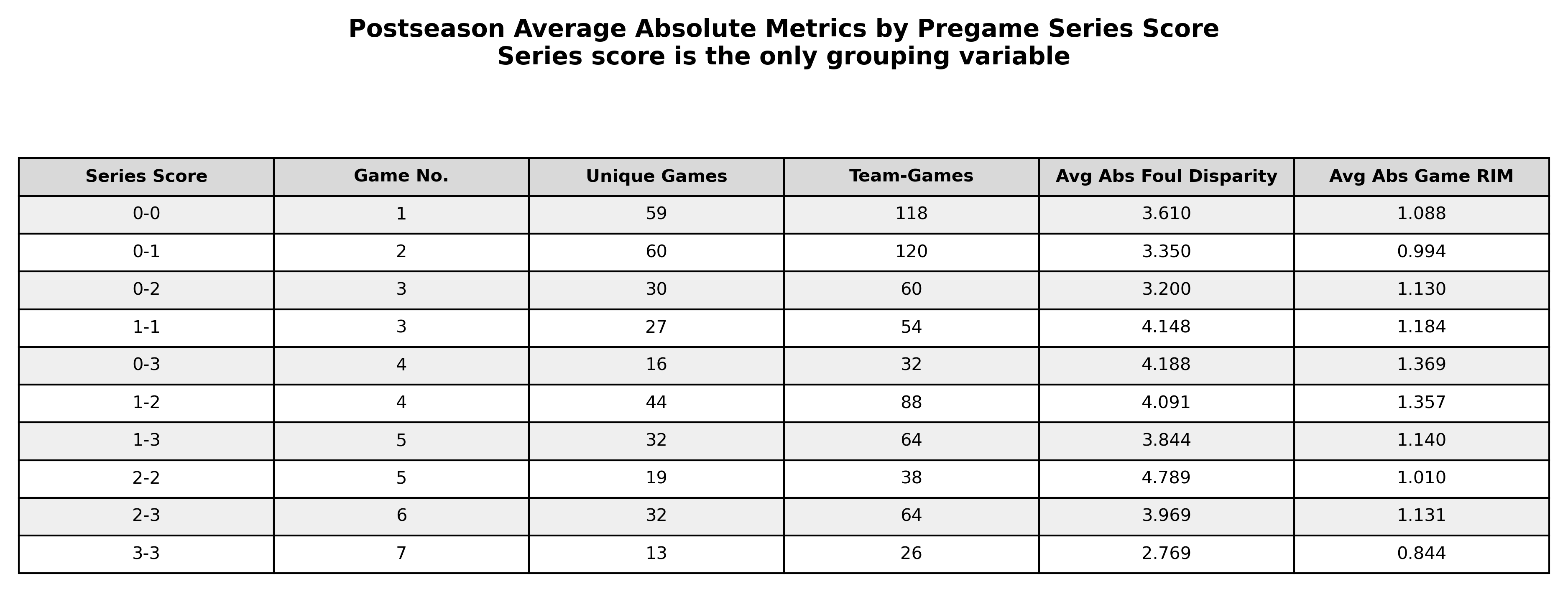}
    \caption{Postseason average absolute foul disparity and average absolute game RIM by normalized pregame series score.}
    \label{fig:series-score-averages}
\end{figure}

The postseason changes the strategic context of each foul because every game carries series-level consequences. Figure~\ref{fig:series-score-averages} summarizes average absolute foul disparity and average absolute game RIM by normalized pregame series score. Mirrored states such as 1--0 and 0--1 are collapsed, so the comparison reflects series pressure rather than home-team orientation.

The postseason RIM distribution uses 356 postseason games. The series-score summaries use the 332 postseason games with complete pregame series-state information, corresponding to 664 team-game observations.

The series-score table shows that postseason leverage does not move in a simple straight line with game number. Average absolute game RIM is higher in several later-series states, including 0--3 and 1--2, while 3--3 is lower in this grouping despite being the most decisive series state. Foul-count imbalance also varies by context, with 2--2 showing the largest average absolute foul disparity.

\begin{figure}[!htbp]
    \centering
    \includegraphics[width=0.98\linewidth]{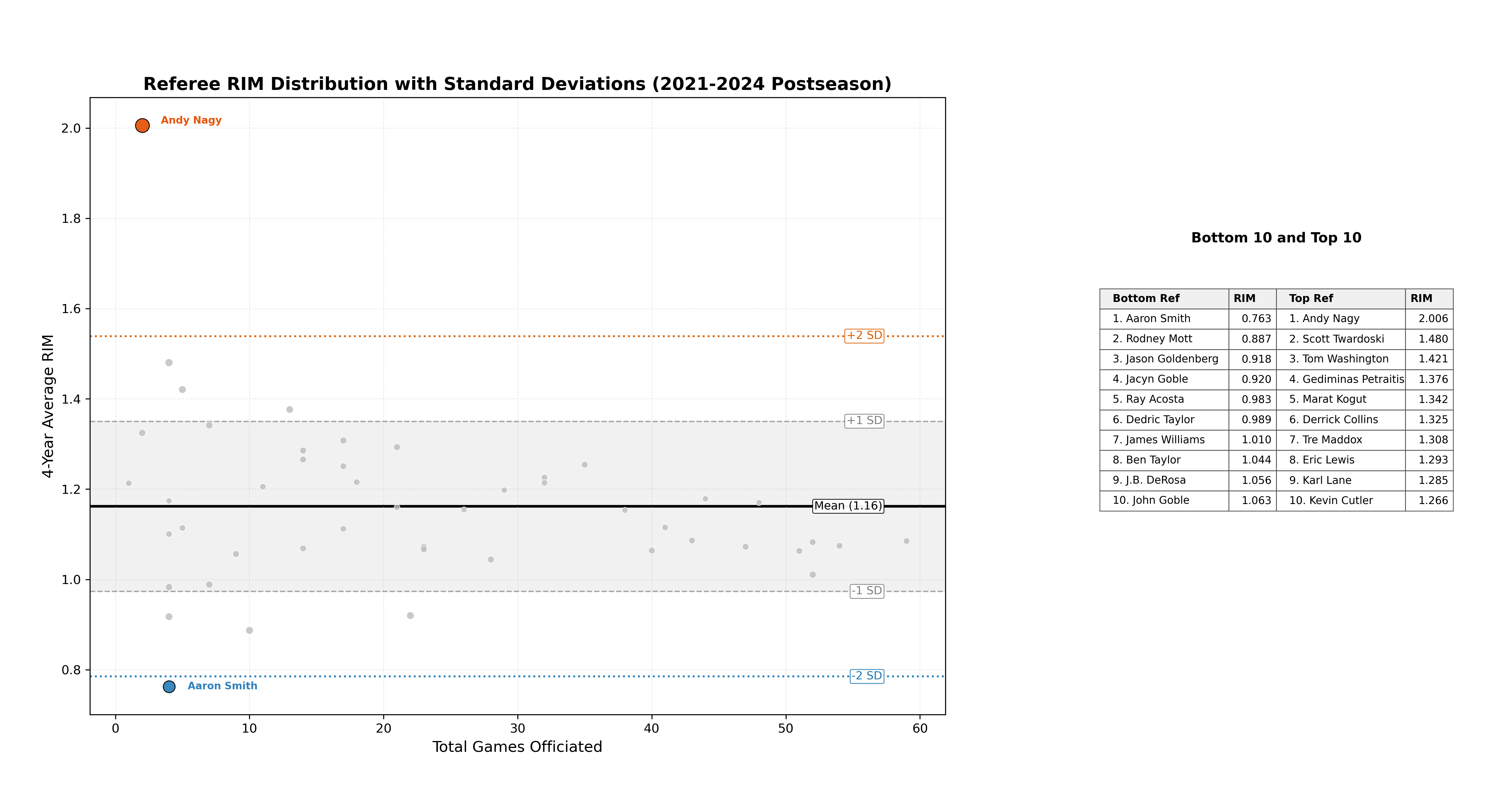}
    \caption{Postseason RIM distribution with standard-deviation bands and the bottom ten and top ten referees by average postseason RIM.}
    \label{fig:postseason-rim-scatter}
\end{figure}

Figure~\ref{fig:postseason-rim-scatter} shows the postseason referee distribution. The sample is necessarily smaller than the regular season, so the estimates should be interpreted with more caution. Still, the figure is useful because it tests whether postseason context produces a different set of high-impact referee-game associations.

Most officials sit in a compact band around the mean, while a small number of points create much of the spread. Andy Nagy appears as the clearest high-side outlier and Aaron Smith appears on the low side, but these should be read as high-leverage associations in the observed postseason sample rather than stable individual ratings.

\FloatBarrier

\section{Home/Away Heterogeneity}
\begin{figure}[H]
    \centering
    \includegraphics[width=0.96\linewidth]{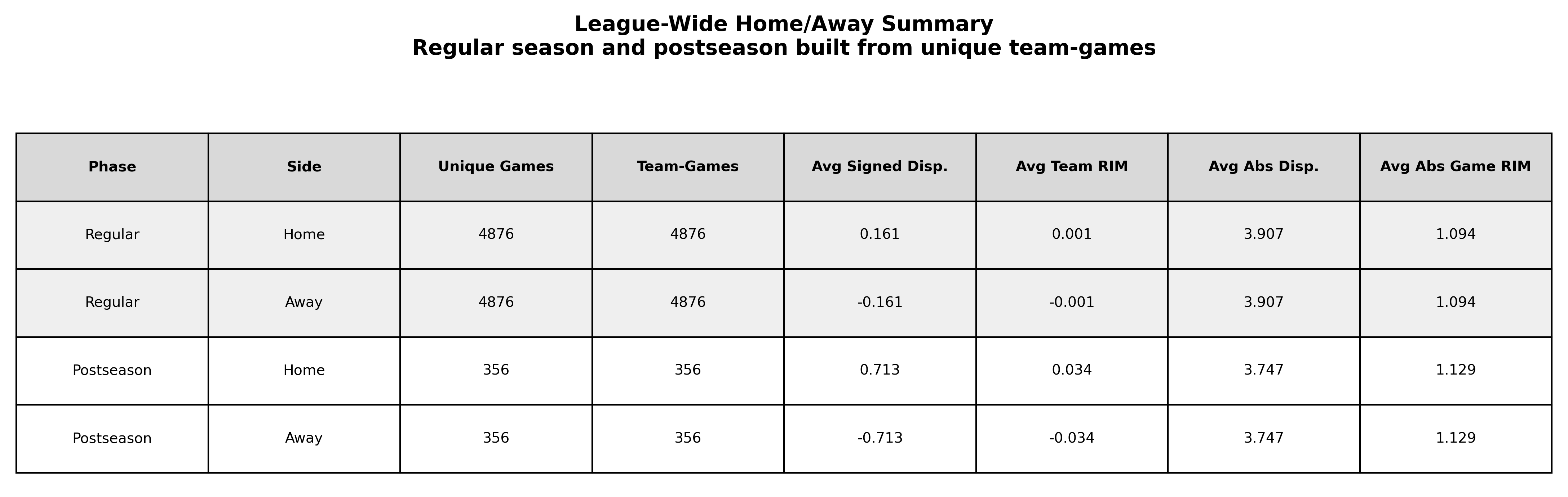}
    \caption{Combined regular-season and postseason home/away summary using signed foul disparity and signed team RIM.}
    \label{fig:home-away-combined}
\end{figure}

Prior work has emphasized home-court pressure and crowd effects in officiating \cite{garicano2005,price2012,gong2022}. Figure~\ref{fig:home-away-combined} compares league-wide home and away outcomes using signed foul disparity and signed team RIM. The regular-season home association is small in aggregate, while the postseason association is larger.

The home/away summaries use 9,752 regular-season team-game observations and 712 postseason team-game observations.

\begin{figure}[H]
    \centering
    \includegraphics[width=0.90\linewidth]{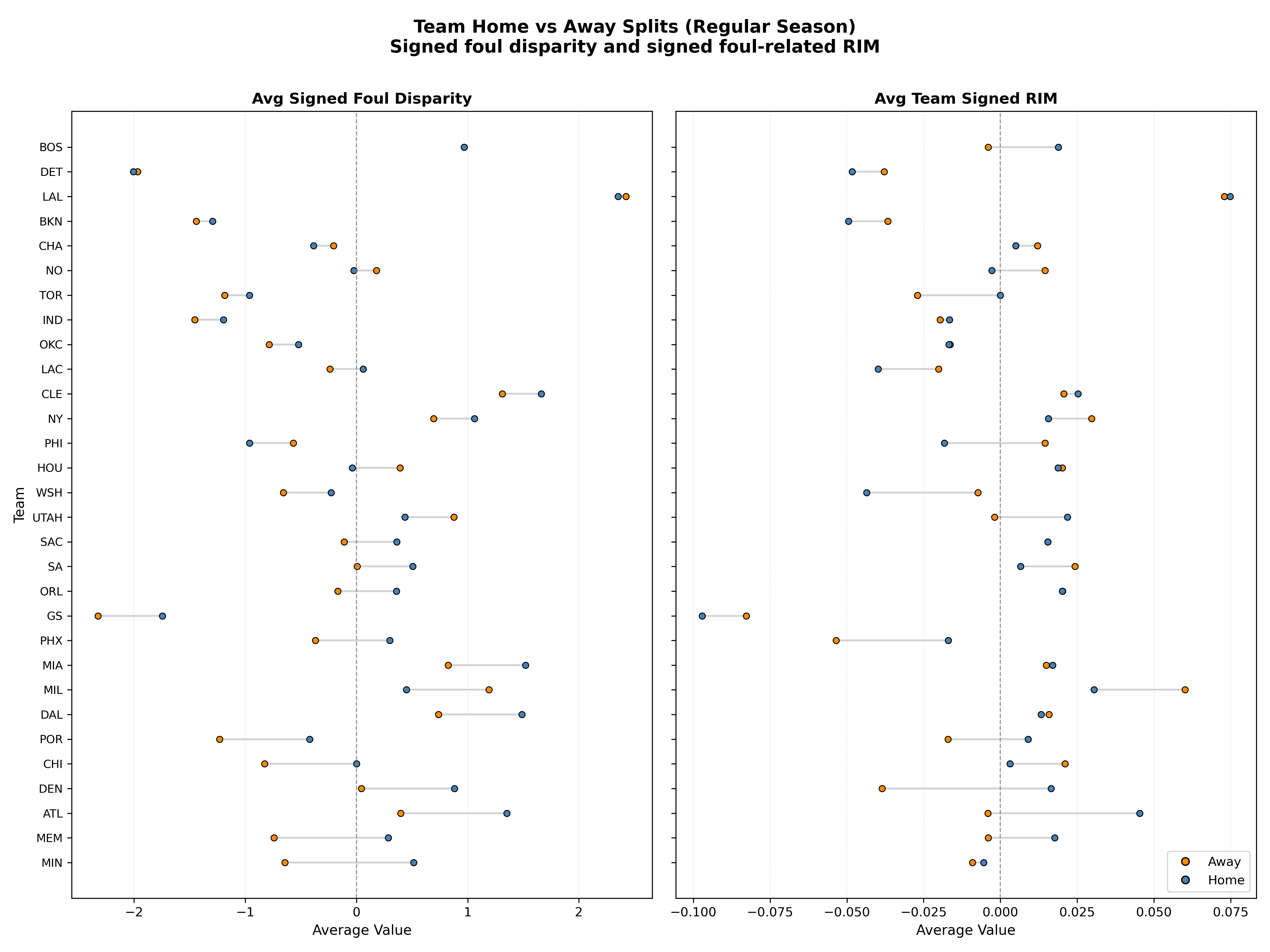}
    \caption{Regular-season team-specific home/away splits. The league average is close to neutral, but individual teams show large positive and negative departures.}
    \label{fig:team-home-away}
\end{figure}

Aggregate neutrality can mask team-specific heterogeneity. Figure~\ref{fig:team-home-away} separates regular-season home/away outcomes by franchise. The Lakers sit high on the positive side, while Golden State is strongly negative, especially away from home. The league-wide average therefore should not be interpreted as evidence that all teams experience similar foul-call environments.

\FloatBarrier

\section{Referee-Team Heterogeneity}
The next question is whether team-level patterns are mirrored in specific referee-team contexts. For each referee-team pair, let \(y_{ij}\) be the observed average outcome for referee \(i\) and team \(j\), let \(a_i\) be the referee average, let \(b_j\) be the team average, and let \(m\) be the global average. The excess value is
\[
x_{ij}=y_{ij}-(a_i+b_j-m).
\]
A positive value means the listed team performs better with that referee than the additive baseline predicts. A negative value means it performs worse.

The regular-season referee-team outlier analysis starts from the expanded panel of 29,256 referee-team-game rows, which corresponds to the same 4,876 unique regular-season games. Applying the five-game minimum leaves 2,175 qualified referee-team pairs.

\begin{figure}[!htbp]
    \centering
    \includegraphics[width=0.82\linewidth]{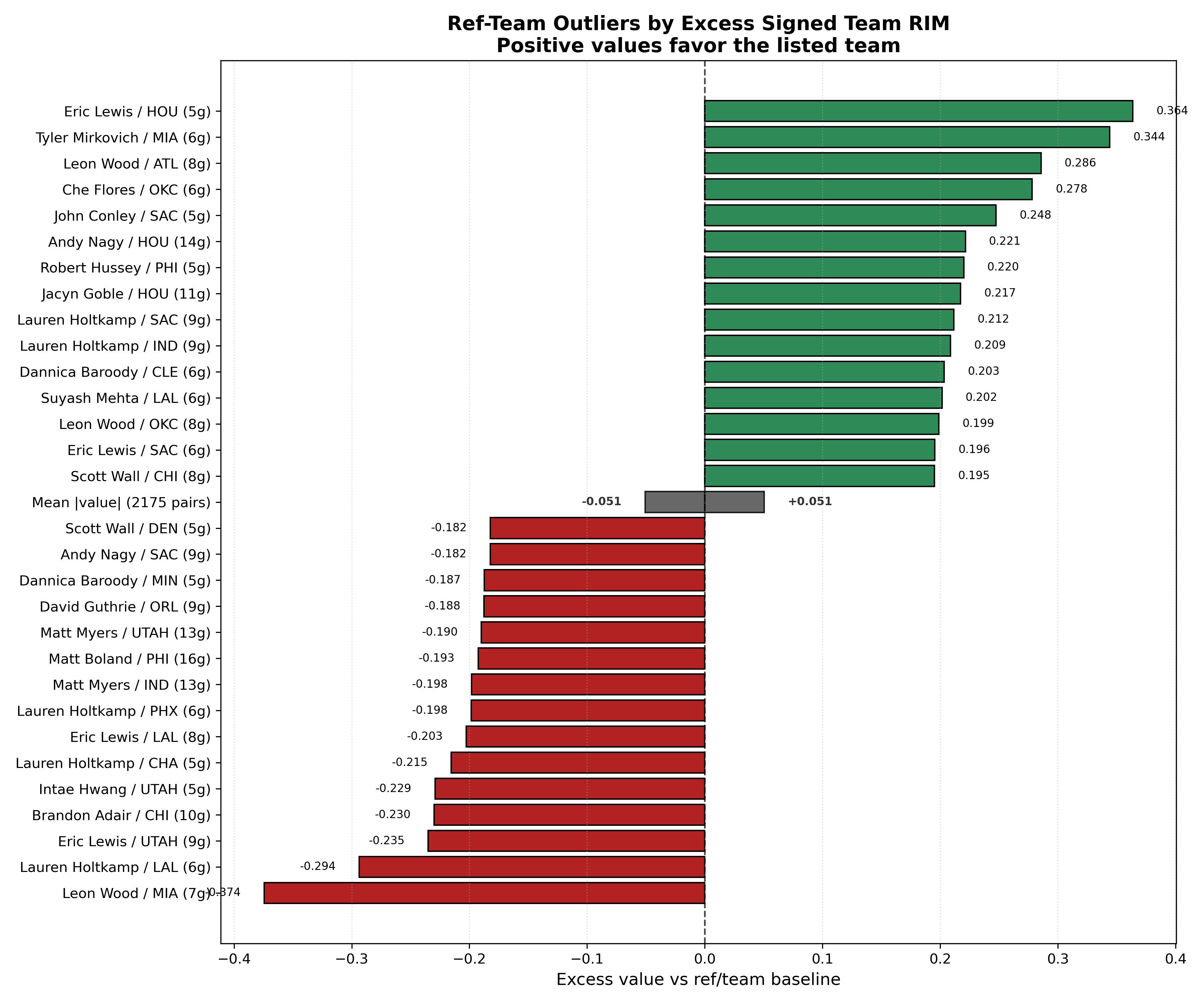}
    \caption{Largest regular-season referee-team outliers by excess signed team RIM. Positive values favor the listed team.}
    \label{fig:ref-team-rim-outliers}
\end{figure}

Figure~\ref{fig:ref-team-rim-outliers} shows the largest excess signed team RIM outliers. These are diagnostic screening results. They indicate where the data depart from an additive referee-plus-team baseline, not whether a referee intentionally favored or harmed a team. Because crew assignment, matchup style, and game script can all contribute to these residuals, the results should be treated as candidates for review rather than as standalone findings.

\begin{figure}[!htbp]
    \centering
    \includegraphics[width=0.82\linewidth]{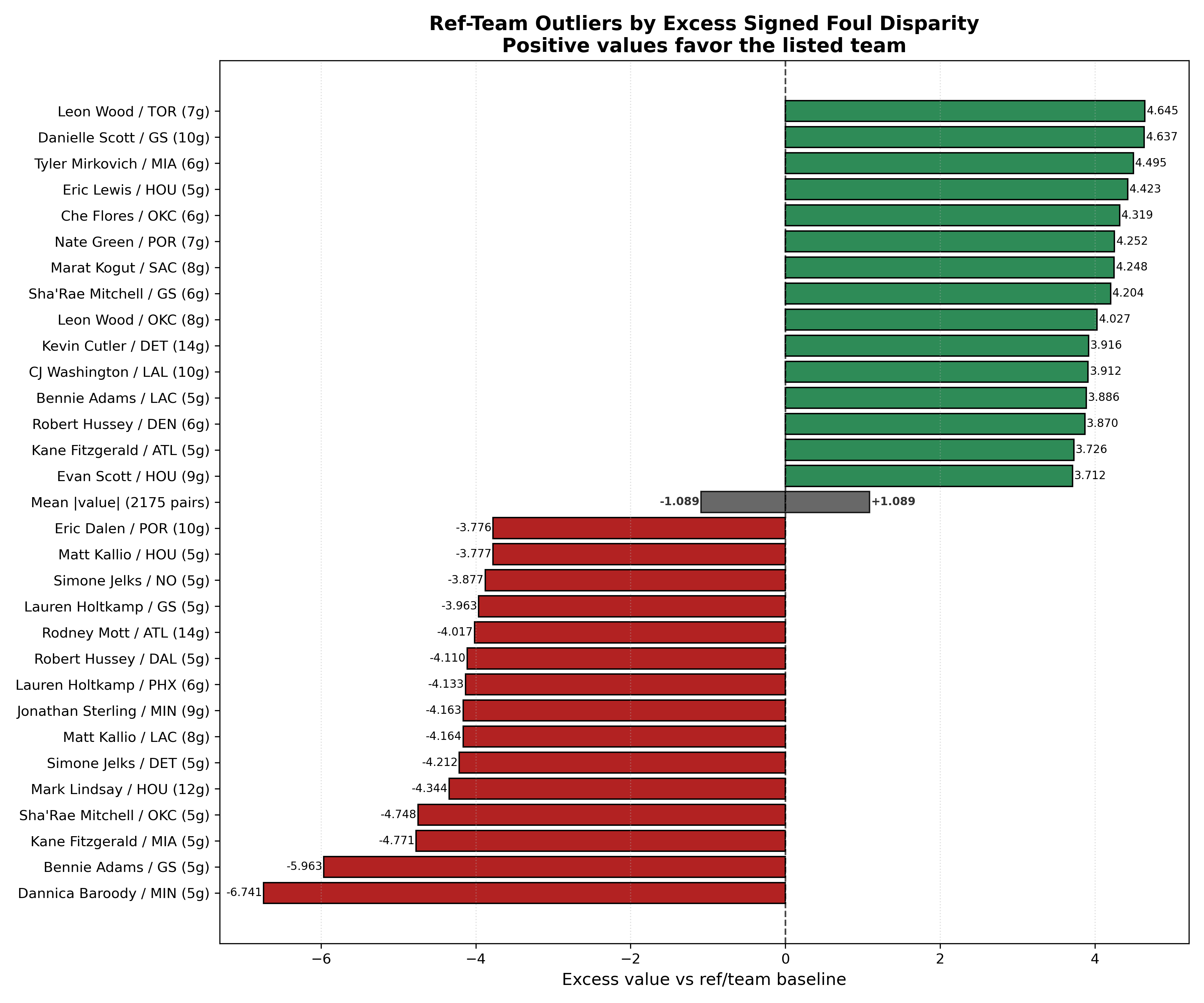}
    \caption{Largest regular-season referee-team outliers by excess signed foul disparity. Positive values favor the listed team.}
    \label{fig:ref-team-disparity-outliers}
\end{figure}

Figure~\ref{fig:ref-team-disparity-outliers} repeats the exercise for signed foul disparity. The comparison is useful because RIM and foul disparity do not always identify the same pairings. Some referee-team combinations look unusual only after conditioning on that specific interaction, which reinforces the value of using both count-based and leverage-weighted summaries.

\FloatBarrier

\section{Controlled Associations and Robustness}
The previous sections are descriptive. In this section, that interpretation will be kept intact. However, we will use regression to serve as a stress test, to observe whether the largest descriptive patterns remain visible after basic conditioning on observable context. If a raw outlier disappears after these controls, it is likely explained by schedule, opponent, home/away setting, season, or playoff state. If it persists, the result is still not causal, but it becomes a stronger candidate for closer review.

The controlled team-side regressions use 9,752 regular-season team-game observations. The series-state regressions use 332 postseason games with complete pregame series-state information. The referee-team residual models use the expanded 29,256-row regular-season referee-team panel, corresponding to 4,876 unique games, and report 2,175 qualified pairings for each metric.

For team-side outcomes, let \(y_{ig}\) denote either signed foul disparity or signed team RIM for team \(i\) in game \(g\). The baseline specification conditions on the most direct observable context available in the dataset:
\[
y_{ig}
=
\alpha
+\beta h_{ig}
+\tau_i
+\omega_{j(i,g)}
+\lambda_t
+\pi_p
+\varepsilon_{ig},
\]
where \(h_{ig}\) indicates home status, \(\tau_i\) are team fixed effects, \(\omega_{j(i,g)}\) are opponent fixed effects, \(\lambda_t\) are season fixed effects, and \(\pi_p\) are playoff series-state controls when applicable. Standard errors are clustered by game.

This specification should be read as a conditioning exercise. Team fixed effects compare a team against its own average environment, opponent fixed effects reduce schedule-composition concerns, season fixed effects absorb league-year shifts, and series-state controls isolate playoff context. The estimates therefore answer a narrower question than causality: after these adjustments, which patterns remain unusually positive or negative?

Robustness is summarized using the strength of an omitted variable needed to explain a remaining estimate away. If the coefficient of interest has \(t\)-statistic \(t\) and residual degrees of freedom \(\nu\), the equal-strength robustness value is
\[
\rho=\frac{-t^2+\sqrt{t^4+4\nu t^2}}{2\nu}.
\]
Here \(\rho\) is the common partial association an omitted variable would need with both the treatment and the outcome, conditional on the included controls, to move the estimate to zero.

\begin{figure}[!htbp]
    \centering
    \includegraphics[width=0.96\linewidth]{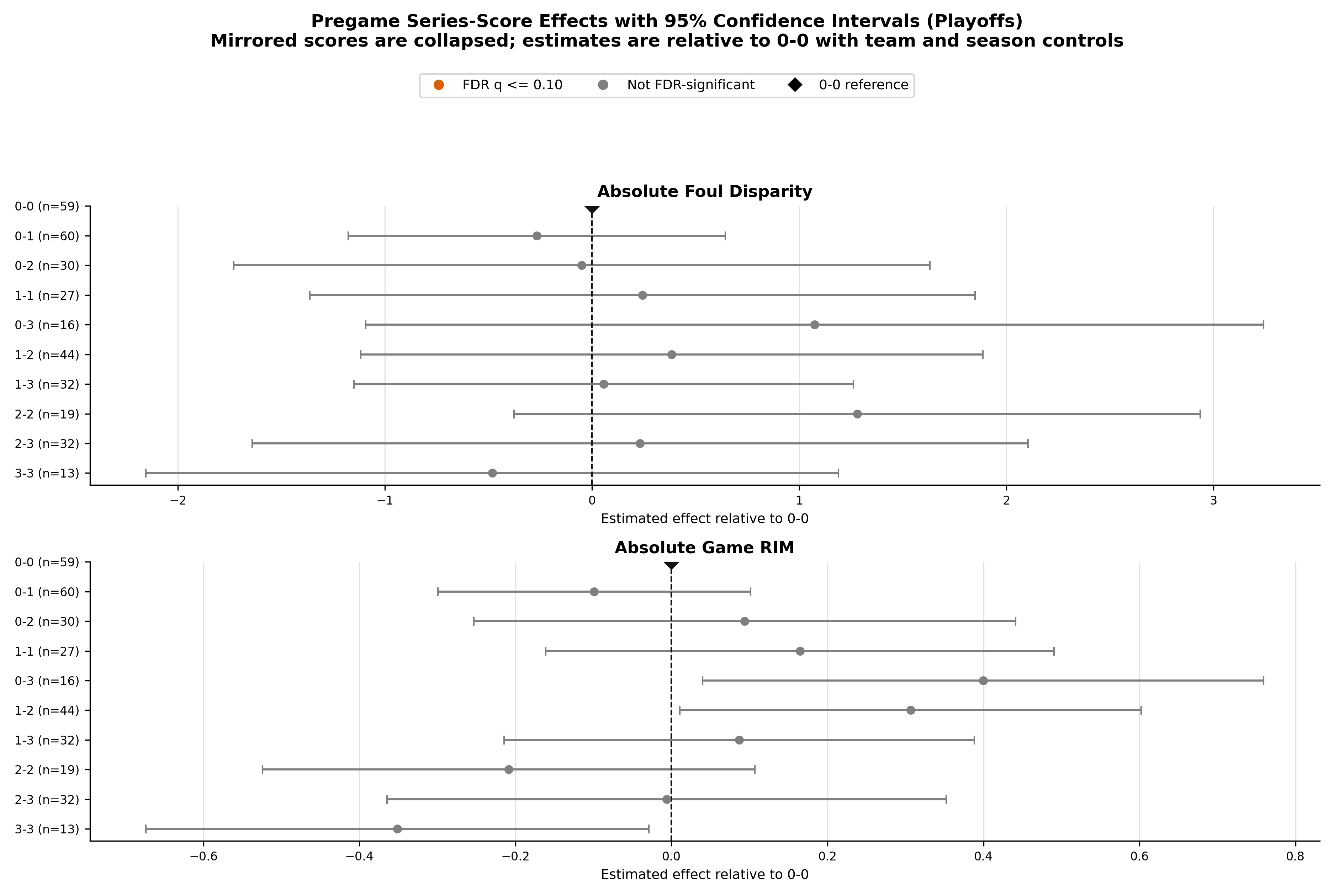}
    \caption{Pregame series-score effects with 95\% confidence intervals in the postseason. Mirrored scores are collapsed and estimates are relative to 0--0.}
    \label{fig:series-score-inference}
\end{figure}

Figure~\ref{fig:series-score-inference} applies the framework to postseason series state. Estimates compare each normalized pregame series score with 0--0 while controlling for home team, away team, and season. The point is not to claim that a series score causes a particular officiating pattern. Rather, the figure asks whether the descriptive series-state differences remain after accounting for which teams are playing, who is home, and which season the game belongs to. The confidence intervals are wide, so the series-score results should be interpreted as suggestive context signals rather than conclusive effects.

\begin{figure}[!htbp]
    \centering
    \includegraphics[width=0.96\linewidth]{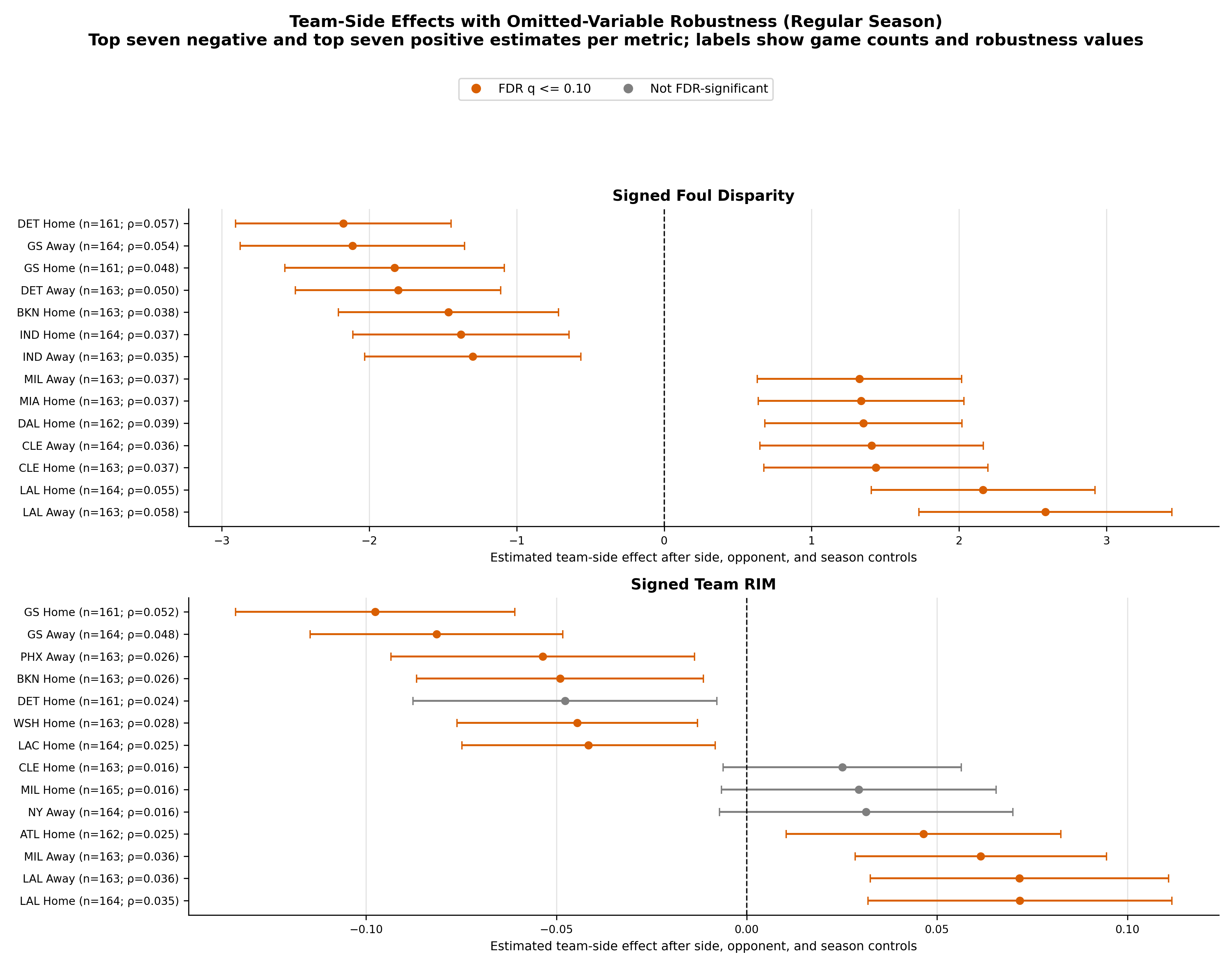}
    \caption{Regular-season team-side effects with omitted-variable robustness. Positive values favor the listed team-side combination.}
    \label{fig:team-side-sensitivity}
\end{figure}

Figure~\ref{fig:team-side-sensitivity} examines whether the largest team-side deviations remain after home-status, opponent, and season controls. The Lakers remain positive and Golden State remains negative on both foul-disparity and signed-RIM dimensions, which means these patterns are not explained away by the most basic observable schedule and season adjustments. At the same time, the robustness values are meaningful but modest. The correct interpretation is therefore not that the model proves preferential treatment. Instead, these team-side patterns survive a first-pass stress test while still leaving play style, matchup profile, defensive aggressiveness, and possession-level shot selection as plausible omitted explanations.

\begin{figure}[!htbp]
    \centering
    \includegraphics[width=0.96\linewidth]{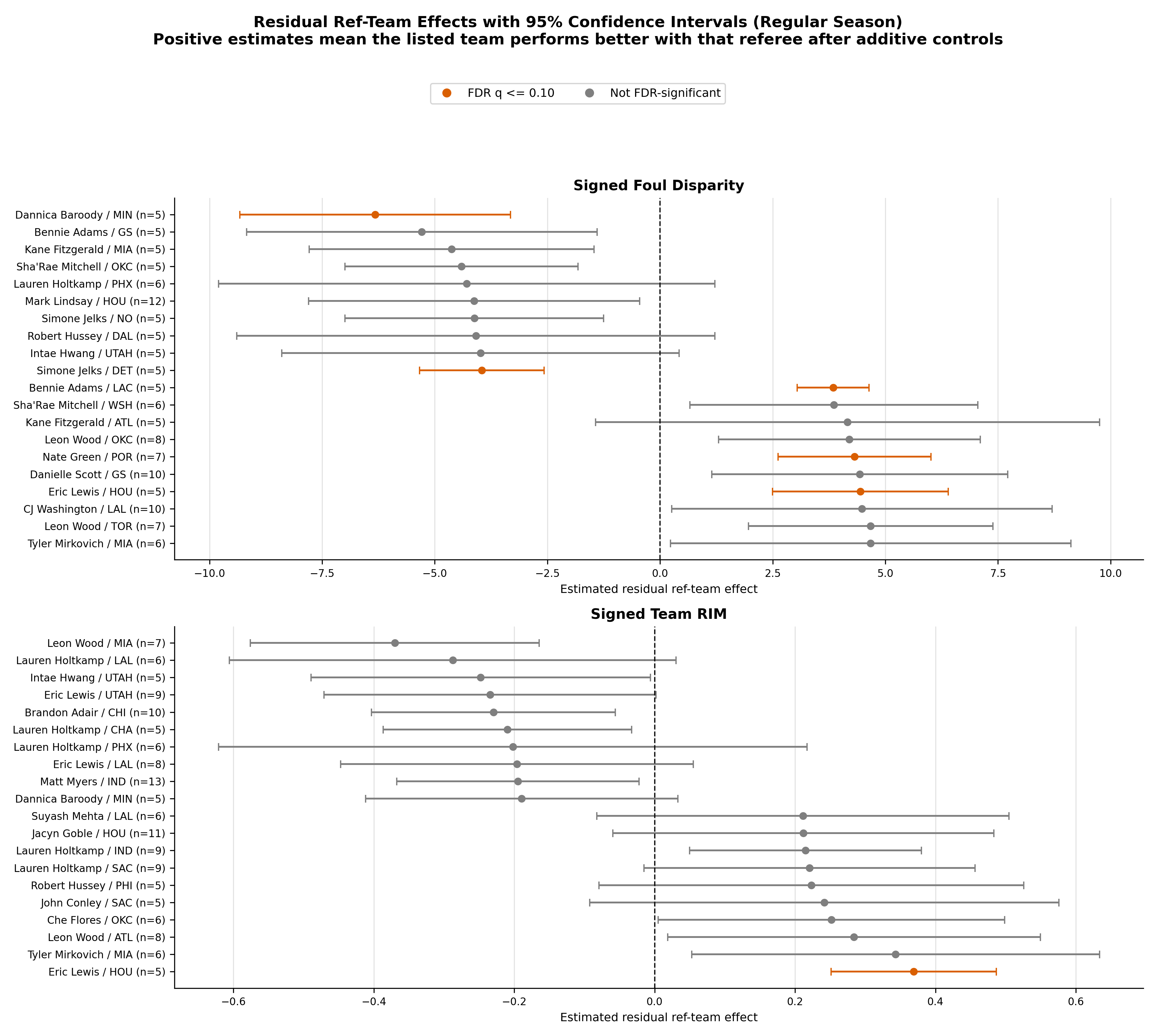}
    \caption{Residual regular-season referee-team effects with 95\% confidence intervals. Positive values mean the listed team performs better with that referee after additive controls.}
    \label{fig:ref-team-inference}
\end{figure}

Figure~\ref{fig:ref-team-inference} presents residual referee-team effects after additive controls. These estimates are the closest analysis in the paper to the referee-team outliers because they ask whether specific pairings remain unusual after subtracting broad referee, team, opponent, and season context. Some pairings continue to show nontrivial residuals, but sample sizes are small for many pairs and intervals are often wide. The figure is therefore best viewed as a prioritization tool for further review rather than as a final causal ranking.

\FloatBarrier

\section{Conclusion}
For decades, the evaluation of NBA officiating has relied heavily on foul counts, whistle rates, and simple free-throw disparities to measure impact. This paper bridges the methodological gap between traditional box-score officiating summaries and the modern deep-context era of basketball analytics. By introducing the Ref Impact Metric (RIM), we provide a reproducible framework that evaluates officiating not merely by the volume of calls made, but by the mathematical leverage those calls exert on the game's expected outcome.

The empirical application of RIM to the 2021--22 through 2024--25 seasons demonstrates that whistle volume and whistle leverage are distinct phenomena. Most regular-season officials cluster near the mean, while officials such as Eric Lewis and Nick Buchert appear at the upper tail of average RIM and others such as Tom Washington appear at the lower tail. Quarter-level decompositions show that high-impact profiles remain elevated across all four quarters rather than being driven solely by fourth-quarter foul accumulation.

Evaluating officiating through a leverage-weighted lens also reveals localized variance that aggregate statistics obscure. While league-wide home/away outcomes appear close to neutral, franchise-specific splits show substantial divergence. Teams such as the Los Angeles Lakers and Golden State Warriors anchor opposite ends of the spectrum, maintaining their respective positive and negative residual effects even after applying team, opponent, and season controls.

This framework is designed to be diagnostic rather than accusatory. As established through the linear controls and omitted-variable sensitivity checks, the observed patterns represent conditional associations within observational data. Because current public play-by-play feeds lack whistle-level attribution, RIM evaluates crew-level variance. It does not control for every unobserved variable, nor does it convert statistical anomalies into definitive proof of individual referee bias or intent. Instead, the metric acts as a macro-level screening tool, identifying which referee-game associations coincided with meaningful expected-outcome movement.

In short, RIM measures impact: the size and direction of win-probability movement associated with foul events. Bias is a separate causal and normative claim requiring evidence about call correctness, comparable game context, and decision-making mechanisms beyond the public data used here.

A structural limitation of this framework is its reliance on ESPN's proprietary win-probability model. Because the exact algorithmic weights that ESPN uses are not public, the model acts as a ``black box'' measurement instrument. Proprietary models may also undergo unannounced version updates, complicating strict long-term reproducibility. However, because the analysis evaluates comparative distributions, any algorithmic biases are applied uniformly across the referee pool, preserving relative outlier status. Macro-level leverage states are also generally robust across different probability models. Even so, future research should explicitly replicate RIM using alternative win-probability models to test whether referee rankings, team-side outliers, and referee-team patterns are stable across modeling choices. Future work can substitute the ESPN feeds used here with open-source, reproducible state-transition models as they become available for the NBA.

The primary contribution of this work is foundational. By moving officiating metrics out of a vacuum and into a win-probability framework, RIM establishes the groundwork for future research. The literature can build on this metric by integrating it with optical tracking data to address the whistle-attribution problem, or by pairing it with exogenous causal instruments to pursue stricter causal identification. Ultimately, evaluating officiating impact requires the same situational context already expected in player evaluation, and RIM provides the mathematical structure needed to make that transition possible.

\FloatBarrier
\clearpage

\appendix
\setcounter{figure}{0}
\renewcommand{\thefigure}{A\arabic{figure}}
\section{Supplementary Figures}
The appendix collects additional diagnostics that support the main text. These figures are not used to introduce new claims. They provide broader versions of the component checks, additional referee-team views, and quarter-level foul-disparity context.

\subsection{Metric Component Checks}
\begin{figure}[!htbp]
    \centering
    \includegraphics[width=0.84\linewidth]{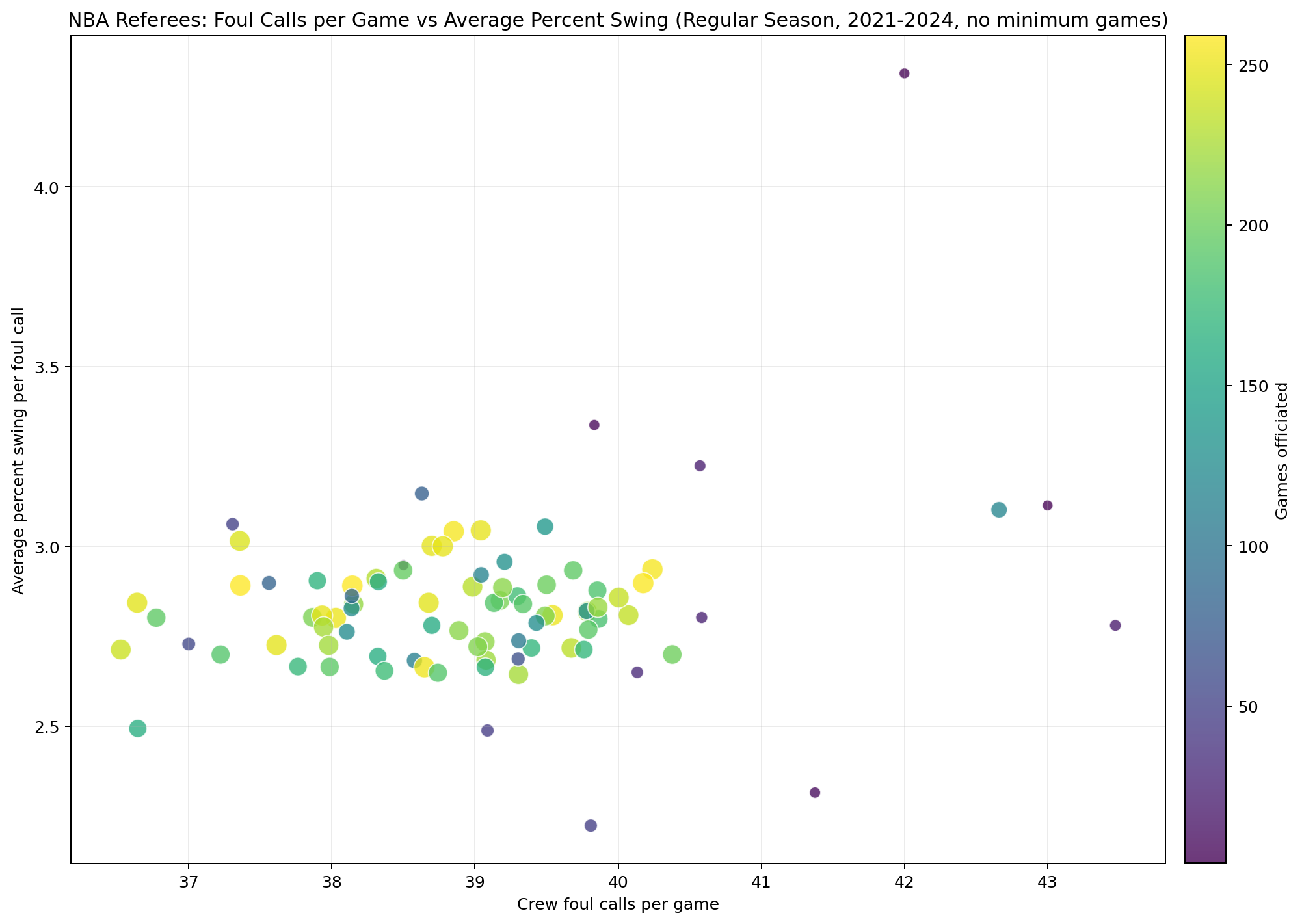}
    \caption{Regular-season foul calls per game plotted against average percent swing per call without a minimum-games threshold.}
    \label{fig:app-calls-vs-swing-no-min}
\end{figure}

Figure~\ref{fig:app-calls-vs-swing-no-min} repeats the foul-volume versus swing-per-call comparison without the 50-game minimum used in the main text. Including the full regular-season referee sample shows that the weak relationship between call volume and call leverage is not only an artifact of the qualified-referee threshold.

\subsection{Quarter-Level Context}
\begin{figure}[!htbp]
    \centering
    \includegraphics[width=0.96\linewidth]{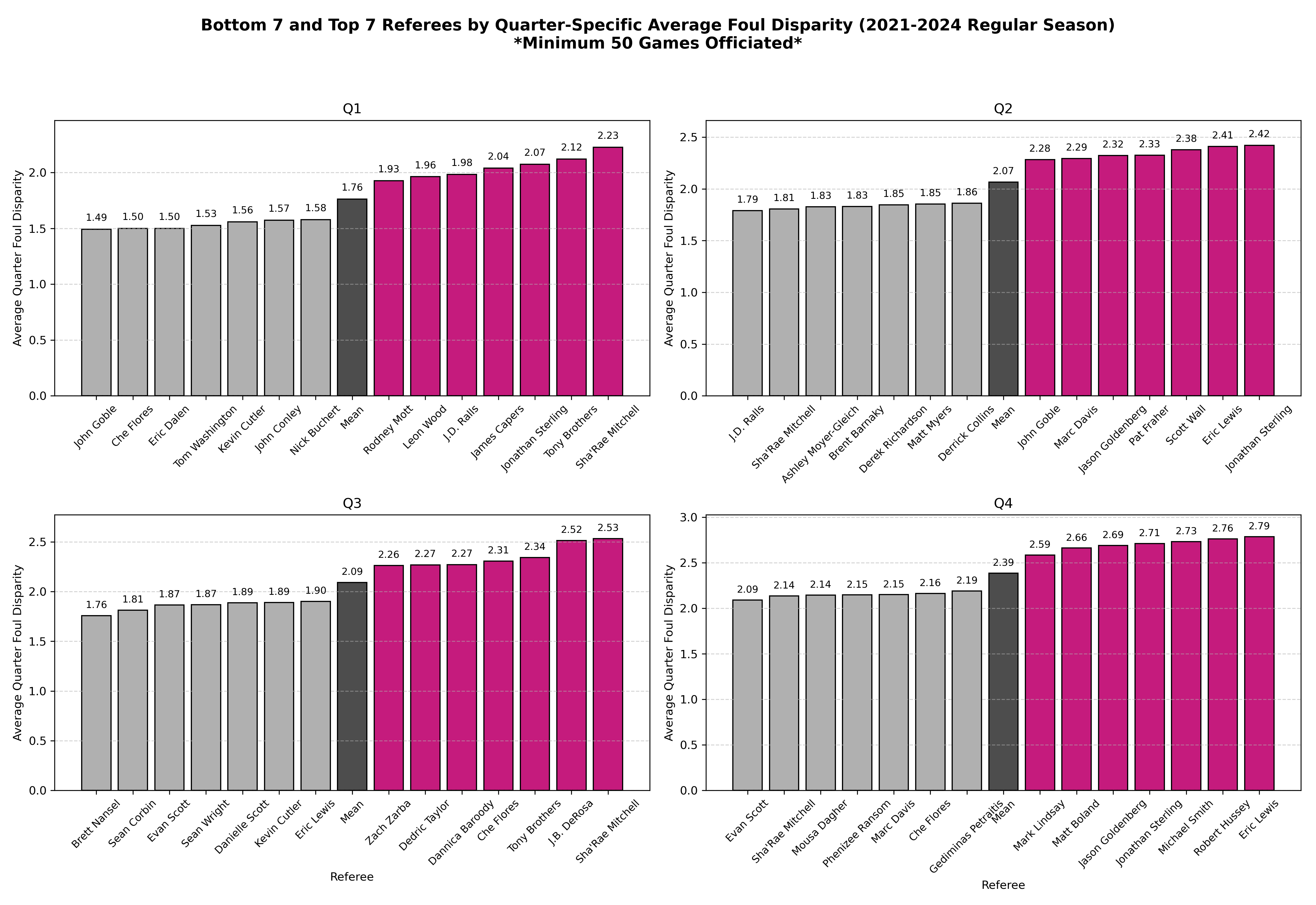}
    \caption{Quarter-specific foul disparity by referee in the regular season.}
    \label{fig:app-quarter-disparity}
\end{figure}

Figure~\ref{fig:app-quarter-disparity} supplements the quarter-specific RIM analysis by showing how foul disparity varies across quarters by referee. This helps separate leverage-weighted patterns from simple foul-count imbalance within game segments.

\subsection{Referee-Team Diagnostics}
\begin{figure}[!htbp]
    \centering
    \includegraphics[width=0.88\linewidth]{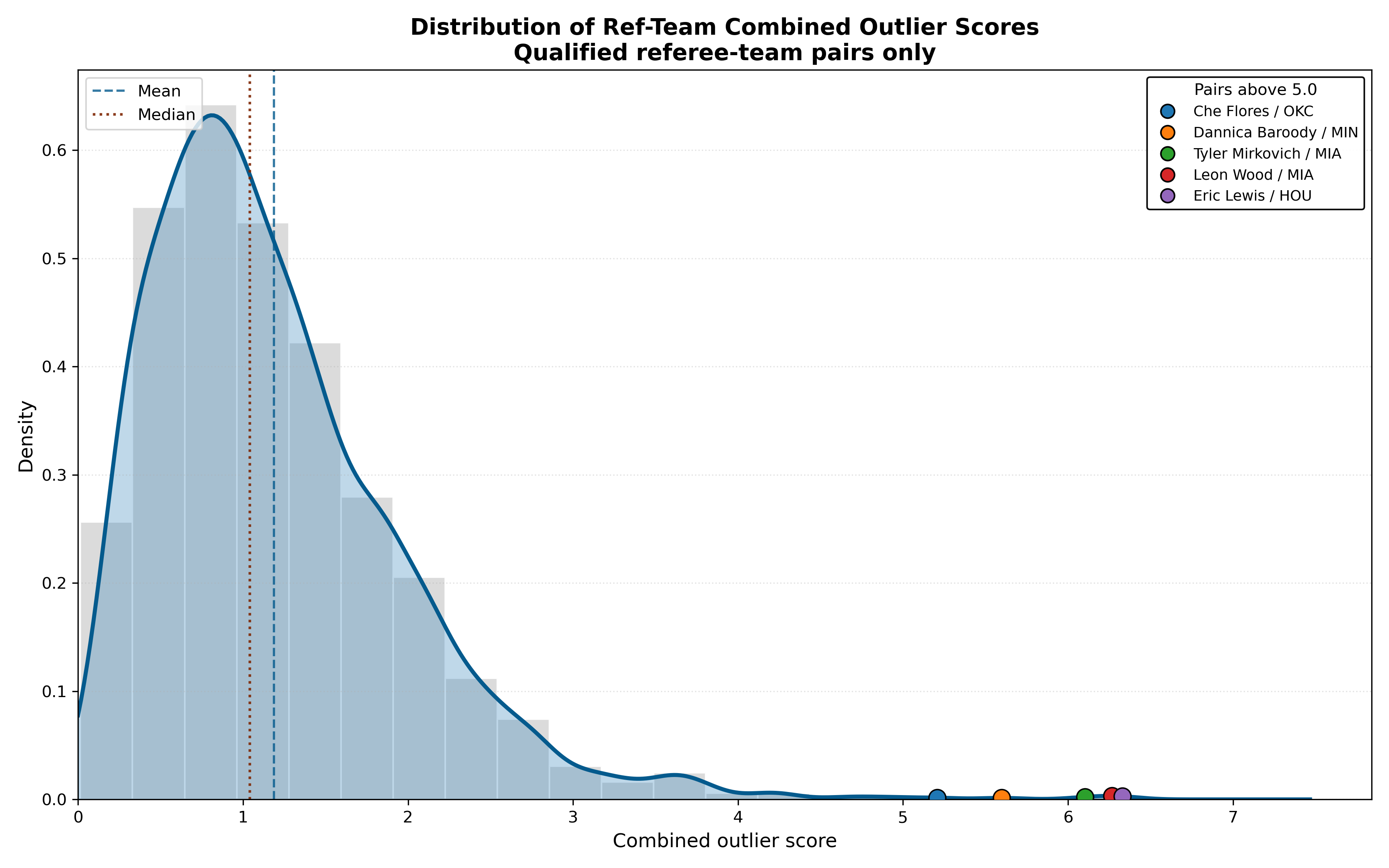}
    \caption{Regular-season referee-team outlier map using excess signed foul disparity and excess signed team RIM. Color indicates the combined z-score from adding the two standardized excess metrics. Positive values favor the listed team.}
    \label{fig:app-ref-team-scatter}
\end{figure}

Figure~\ref{fig:app-ref-team-scatter} shows the referee-team interaction space using excess signed foul disparity and excess signed team RIM. The color scale uses a combined z-score, formed by adding the z-score for excess signed foul disparity and the z-score for excess signed team RIM. It provides a two-dimensional view of the outlier tables in the main text.

\begin{figure}[!htbp]
    \centering
    \includegraphics[width=0.84\linewidth]{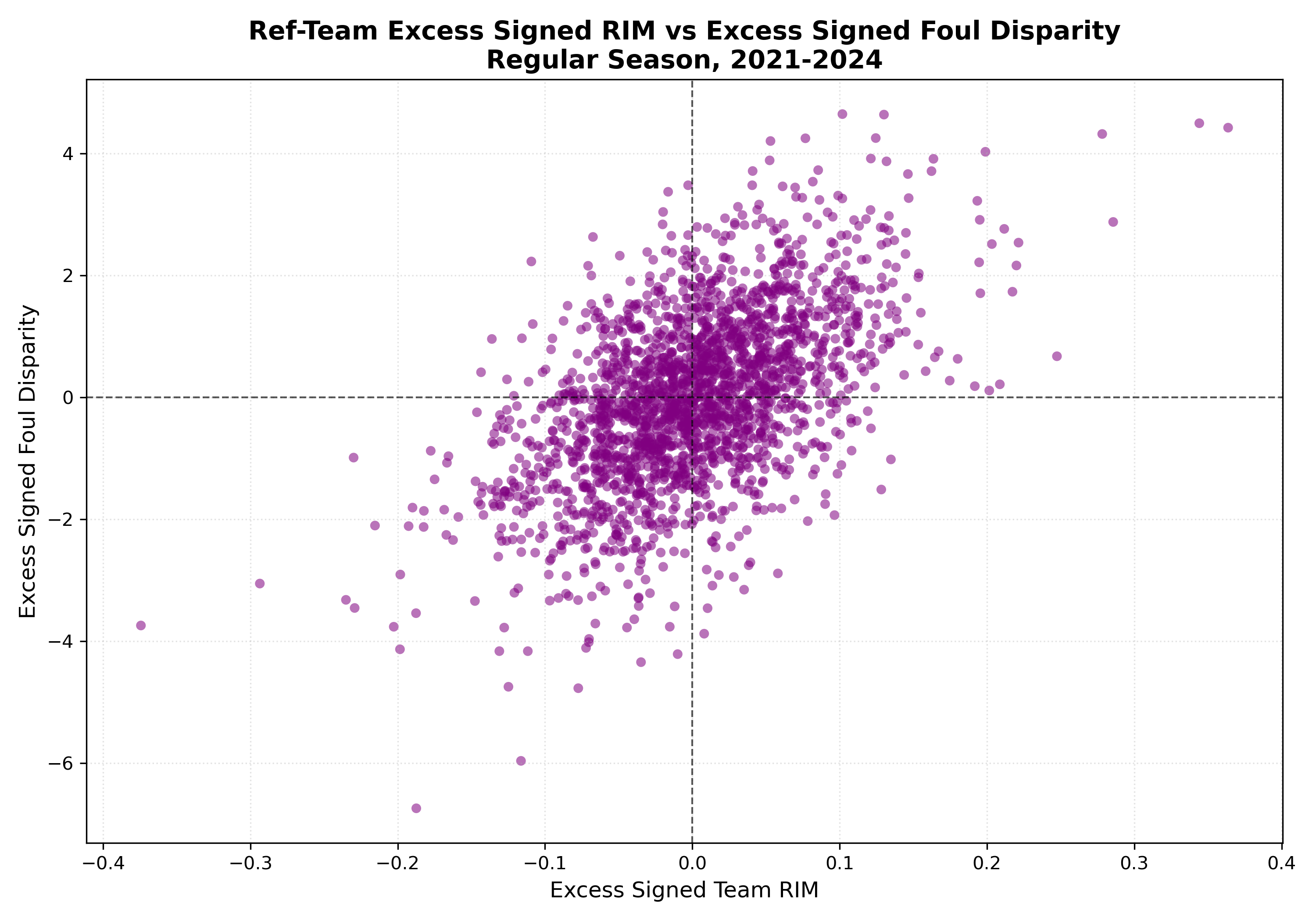}
    \caption{Regular-season referee-team excess signed team RIM plotted against excess signed foul disparity. The two measures are correlated overall, with exceptions where foul-count imbalance and leverage-weighted movement diverge.}
    \label{fig:app-excess-rim-disparity}
\end{figure}

Figure~\ref{fig:app-excess-rim-disparity} compares excess signed team RIM against excess signed foul disparity across referee-team observations. The figure confirms that signed foul disparity and signed RIM are correlated overall, but it also shows exceptions where count imbalance and leverage-weighted win-probability movement diverge. This supports the main-text claim that leverage-weighted and count-based summaries identify related but non-identical patterns.

\FloatBarrier
\clearpage

\bibliographystyle{unsrt}
\bibliography{nba_officiating_refs}

\end{document}